\documentclass[final]{eps}

\def\xhat{{\bf \hat x}}
\def\yhat{{\bf \hat y}}
\def\zhat{{\bf \hat z}}

\newcommand{\be}{\begin{equation}}
\newcommand{\ee}{\end{equation}}

\usepackage{graphicx}
\usepackage{times}

\volumenumber{61}
\publishyear{2009}
\frompage{\pageref{firstpage}}
\topage{\pageref{finalpage}}

\shorttitle{M. G. Linton \lowercase{\textit{et al.}}: 
Patchy Reconnection in a Y-Type Current Sheet}

\title{Patchy Reconnection in a Y-Type Current Sheet}
\author{M. G. Linton$^1$, C. R. DeVore$^1$, and D. W. Longcope$^2$}
\affiliation{$^1$Naval Research Laboratory, Washington, DC\\
             $^2$Montana State University, Bozeman, MT}

\received{Nov. 21, 2007}\revised{Mar. 24, 2008}\accepted{Jun. 20, 2008}

\abstract{
We study the evolution of the magnetic field in a Y-type current sheet
subject to a brief, localized magnetic reconnection event. 
The reconnection produces up- and down-flowing reconnected
flux tubes which rapidly decelerate when they hit the
Y-lines and underlying magnetic arcade loops at the ends of the 
current sheet. This localized reconnection outflow followed by a rapid 
deceleration reproduces the observed behavior of post-CME 
downflowing coronal voids. These simulations support the hypothesis that
these observed coronal downflows are the retraction of magnetic
fields reconnected in localized patches in the high corona.}


\begin{document}
\label{firstpage}
\maketitle
\copyrighttext{}

\section{Introduction}

Reconnection is believed to be a key process allowing the excitation
of solar flares and coronal mass ejections (CMEs). The reconnection releases
significant magnetic energy, leading to solar flare heating, and changes 
magnetic topologies, allowing CME magnetic fields to erupt and 
escape the solar corona into interplanetary space.  Observations of 
the flaring which occurs behind recently erupted CMEs show 
downflowing voids (see, e.g., McKenzie \& Hudson 1999, Gallagher et al.\ 2002,
Innes et al.\ 2003, Asai et al.\ 2004, Sheeley et al.\ 2004) which 
push their way through the heated flare plasma in the high corona.
These downflowing voids, observed by TRACE, Yohkoh SXT, 
and Hinode XRT, have been shown to be evacuated structures. 
They are therefore not cool, dense plasma blobs being pulled 
down by gravity, but rather appear to be be evacuated loops of magnetic
field being pulled down by the magnetic tension force (McKenzie \& Hudson 
1999). The three dimensional (3D) structure of these voids breaks up the two 
dimensional (2D) symmetry of the flare current sheet and arcade, implying that
the reconnection which creates these voids occurs in localized 3D
patches rather than uniformly along the current sheet. The
reconnected field from this patchy reconnection takes the shape of 
individual 3D flux tubes rather than extended 2D sheets of field.  
In this letter, we study whether the formation of magnetic loops 
high in the corona via a 3D patch of reconnection creates structures 
consistent with the morphology and dynamics of these coronal voids.

In Linton \& Longcope (2006), we showed that the shapes
and evolutions of such magnetic loops in a one dimensional 
(1D) Harris type current sheet are consistent with observations of
these downflowing loops. The cross sections of the simulated magnetic
loops form teardrop shapes, similar to that of the voids, while
the 3D structure of the loops is similar to the structure of the
coronal loops which appear below these voids, e.g., as seen by Sheeley
et al.\ (2004). 

\begin{figure*}[t]
\centerline{\includegraphics[width=16.0cm,clip]{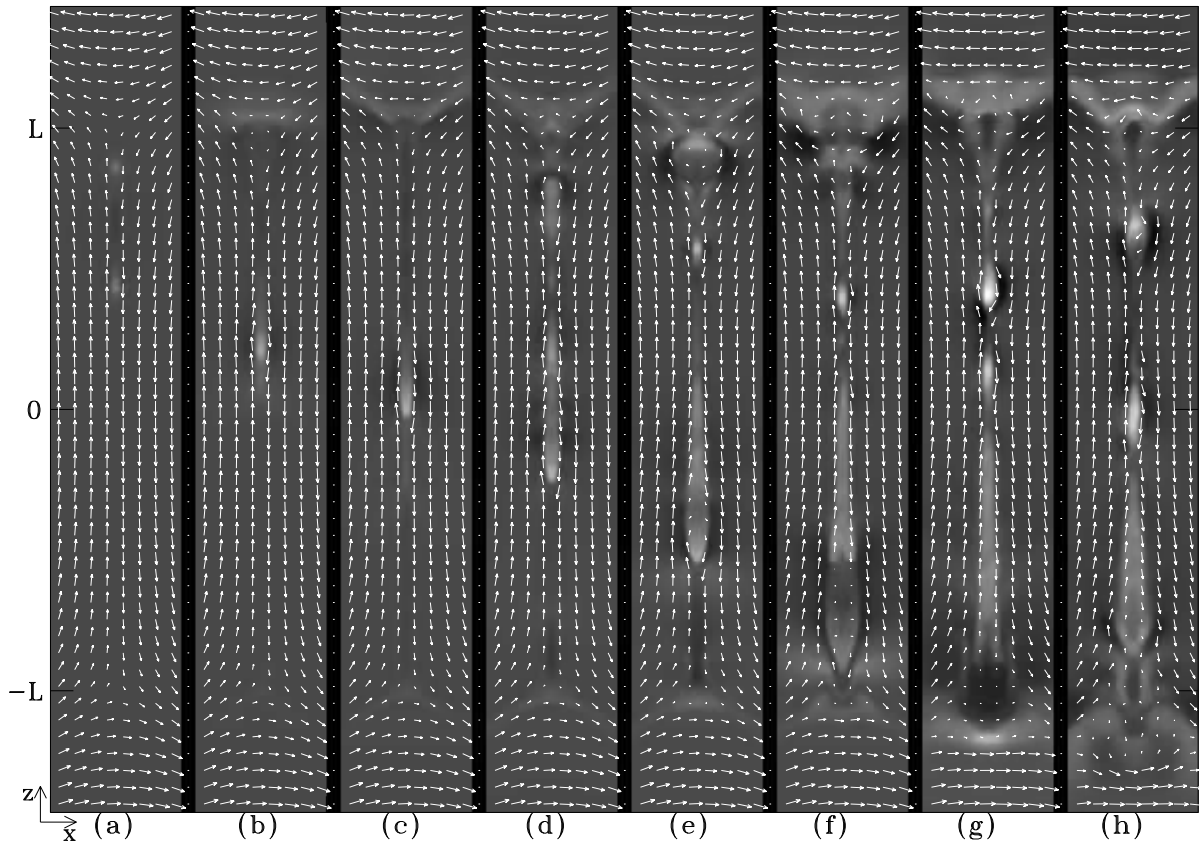}} 
\caption{Magnetic field of the Y-type current sheet in the $y=0$ plane,
with a magnetic reconnection patch imposed at $z=2L/3$ for a
time $tv_{A0}/L=0.6$.  The vectors show
the magnetic field in the plane, while the greyscale shows the 
guide field component, with white representing maximum positive field. 
The panels show the simulation at times 
$tv_{A0}/L=[0.2,0.7,1.1,1.6,2.1,2.6,3.1,3.5]$.
The $x$ boundaries of each panel shown are at $x=\pm 3L/4\pi$.
}
\label{fig:bvect}
\end{figure*}

However, there is a key aspect of the void dynamics which the 
initial 1D nature of the Harris current sheet cannot 
reproduce. This is the rapid deceleration
of the voids once they reach the post-flare arcade loops in the
low corona.  Sheeley et al.\ (2004) have shown that the speeds of these 
voids through the high corona are relatively constant until
they hit the coronal arcade, when they rapidly decelerate.
As the Harris current sheet continues unchanged to the edge of the
simulation, there is no arcade of loops at the base of such 
a current sheet with which the voids collide.  To study this deceleration, 
we therefore now simulate this patchy reconnection in a Y-type current 
sheet (Green 1965). The Y-type current sheet terminates at a
set of magnetic arcade loops, as shown in Figures 1(a) and 2(a).
The intersections of the current sheet and the
outermost of these arcade loops forms the two Y-lines, at $z=\pm L$ 
in Figure 1(a). These arcades make the current sheet more representative 
of a post-CME coronal current sheet with underlying arcade fields.
Note that these Y-lines are not Y-type nulls, as there is a uniform
guide field in both the current
sheet and the arcade, so the field strength does not go to zero
at the Y-lines, even though the reconnection component of the field
does go to zero here.

We study the effect of a localized reconnection event in this
current sheet, focusing on the form of the reconnected field, 
and on whether it decelerates once it hits
the Y-line and the coronal arcade below it.
The current sheet configuration and the simulation setup are
discussed in \S 2, the results are described in \S 3, and
our conclusions are summarized in \S 4.

\section{Simulations}

The simulations were performed using the magnetohydrodynamic (MHD) 
code ARMS (Adaptively Refined MHD Solver) on the
Cray XD1 supercomputer at the Naval Research Laboratory.
The code was used to solve the resistive MHD equations. See
Welsch et al.\ (2005) for a discussion of the code, and a
presentation of the equivalent ideal MHD equations. 
We include an explicit resistivity $\eta$
in the induction equation, 
$\partial{\bf B}/\partial t = \nabla\times({\bf v}\times {\bf B} -
\eta \nabla\times{\bf B})$, and an ohmic heating term in 
the energy equation.

The Y-type magnetic field ($e.g.$, Priest \& Forbes 2000) is 
\be
  B_x+i B_z = -B_0\sqrt{\omega^2/L^2-1},
\ee
where $i$ is the positive value of $\sqrt{-1}$, $\omega\equiv z+ix$, and $L$
is the current sheet half-length.  The guide field is uniform at $B_y=B_0/\pi$.
Here $B_z(z=0, x\sim0)=B_0=44$ in units where the pressure is $p_0=20/3$.
Due to the guide field, the magnetic fieldlines on 
either side of the current sheet form a half angle 
$\zeta=\arctan(B_z/B_y)\sim 2\pi/5$. 
The reconnection field strength $|B_z|$ decreases as $\sqrt{1-z^2/L^2}$ along 
the current sheet, going to zero at the Y-lines at $z=\pm L, x=0$. 
This magnetic configuration is force-free, so the density and gas pressure 
are initially set to be uniform at $\rho=\rho_0=1/2$ and $p=p_0$, 
which sets the ratio of plasma to magnetic pressure near the center of the 
current sheet at $\beta\equiv 8\pi p/|{\bf B}|^2 \sim 0.08$.
The Alfv\'en speed used for normalization,
$v_{A0}=|{\bf B}|/\sqrt{4\pi\rho}$, is measured near 
the center of the current sheet at $z=0$, $x\sim0$.

Extrapolative, zero gradient, open boundary conditions are imposed in the 
$\zhat$ and $\xhat$ directions, while periodic boundary conditions are
imposed in the $\yhat$ direction.  The computational mesh is adaptively 
refined in areas of high current magnitude. This gives a resolution 
ranging from $64$ to $512$ cells in the $\yhat$, and $\zhat$ directions
[-1.9L,1.9L], and a resolution ranging from $32$ to $256$ cells in the 
$\xhat$ direction [-.95L,.95L]. The current sheet is at the highest resolution,
and it is effectively one cell wide, so its thickness is $l=0.007L$.

\begin{figure*}[t]
\centerline{\includegraphics[width=16.0cm,clip]{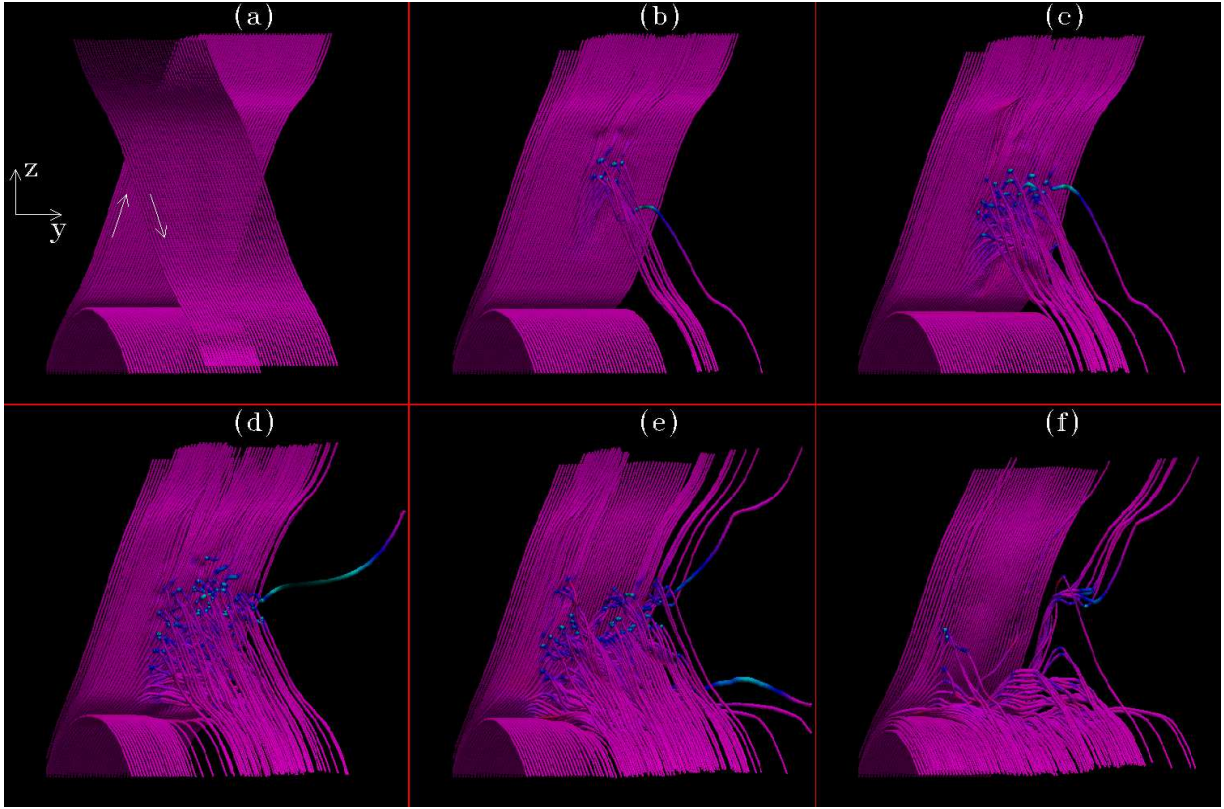}} 
\caption{Fieldlines of the reconnecting Y-type current sheet.
The blue color shows where the electric current parallel to the 
magnetic field is strong.  The panels show the simulation at 
times $tv_{A0}/L=[0.0,1.3,2.1,2.8,3.4,3.9]$. Panel (a) shows fieldlines
traced from both the front and back side of the current sheet, at the
bottom boundary. The rest of the panels show only fieldlines traced
from the back side of the current sheet, at the bottom boundary.
}
\label{fig:fline}
\end{figure*}

The simulation is run with a uniform background resistivity  
$\eta_0$.  To initiate the reconnection, we impose a 
sphere of enhanced resistivity on the current sheet
for the first $tv_{A0}/L=0.6$ of the simulation.  
This resistive enhancement has the form
$ \eta=\eta_0\left(1+99 e^{-r^2/\delta^2}\right)$,
for $r=\sqrt{x^2+y^2+(z-2L/3)^2} < 2\delta$, with $\delta/L=0.087$.
The Lundquist number of the background resistivity is
$S_{\eta}\equiv \delta v_{A0}/\eta_0 = 5000$,
while for the peak resistivity at the center of the reconnection region
$S_{\eta}=50$.  Note that Ugai (2007) have found that reconnection induced 
by locally enhanced resistivity is inhibited if $\delta < 4 l$. As $\delta\sim 10 l$ 
here, the reconnection should not be inhibited by this mechanism.

\section{Results}

The effects of the magnetic reconnection event on the
Y-type current sheet are shown in Figures 1 and 2. Figure 1 shows
a vector magnetogram view of the field in a plane perpendicular to
the current sheet at $y=0$. Figure 1(a) is taken soon after the reconnection event is 
turned on.  The guide field, shown by the greyscale,
is enhanced in two small spots on either side of the reconnection 
event at $z=2L/3$.
These spots are the cross sections of the two recently reconnected
flux tubes flowing away from the reconnection site. 
The guide field is enhanced in these reconnected flux tubes
because the reconnection component of the magnetic field has been annihilated, 
and the guide field must increase to make up for the lost magnetic pressure
(see Hesse et al.\ 1996). 

Figure 2(a) shows the fieldlines on either side of this current sheet 
just before the reconnection starts. Three topologically distinct
sets of fieldlines are shown. One set runs just behind the current sheet
from the bottom to the top of the panel, while a second set
runs in front of the current sheet from the top to the bottom
of the panel, as marked by the arrows. The third set arches
below the current sheet, running from the back of the simulation to
the front, and coming close to the Y-line and the base of the current
sheet at its apex. All three sets of fieldlines are canted
gradually from left to right: this is the effect of the
uniform guide field in the $\yhat$ direction. 
The arched set of fieldlines represents the post-CME arcade loops
while the two other sets of fieldlines represent the coronal current sheet
fieldlines, which may still be connected to the CME
above the second Y-line at the top of the simulation.
For the remaining panels of Figure 2, only the first and third sets
of these fieldlines are traced, so that the second set, which lies
in front of the current sheet, does not obscure the dynamics. 
Both of these sets of fieldlines are traced from the bottom boundary, 
behind the current sheet and arcade. Therefore,
any section of a fieldline which appears in front of the current
sheet above the arcade fieldlines is purely due to reconnection
of front-side fieldlines with the back-side fieldlines.

The reconnection event was initiated very close to the 
upper Y-line, so the upflowing reconnected field quickly
hits the Y-line, in Figure 1(b). The corresponding upward retracting
fieldlines are not shown in Figure 2, since they are not connected
to the lower boundary from which the fieldlines are traced. 
The upflow halts when it hits
the upper Y-line and the arcade fieldlines lying above it. This is the
equivalent of the upgoing fieldlines hitting and merging into the 
recently erupted CME, though this would only happen if the CME were erupting
at a slower rate than the fieldlines retract.

The downflowing part of the reconnected flux is displayed in Figure 
2(b) as the fieldlines which trace upwards with all the field
on the back side of the current sheet, but then suddenly take
a hairpin turn and trace back down to the bottom boundary
on the front side of the current sheet. The blue color of the
fieldlines at these hairpin turns shows that the parallel electric current 
is enhanced and therefore reconnection is strong there
(see, $e.g.$, Schindler et al.\ 1988).
The high placement of the initial reconnection region allows the
cross section of the flux tube carrying this downflowing reconnected flux
to fully take shape. Figures 1(b)-1(f)
show how the cross section of the tube forms into an oblong shape,
reminiscent of the coronal void observations, just
as it did in the Harris sheet experiments of Linton \& Longcope (2006). 
Note that the area of the downflowing flux tube cross section 
continues to grow as it absorbs more reconnected flux, 
even though the enhanced reconnection spot was turned off at 
$tv_{A0}/L=0.6$, just before Figure 1(b). Apparently the disturbance 
caused by the initial reconnection event spontaneously excites subsequent 
patchy reconnection events. 

When this downflowing flux tube hits the Y-line and the arcade field
below it, in Figure 1(f), it decelerates and compresses the arcade 
field, in Figure 1(g), and eventually joins the arcade,
in Figure 1(h). The reconnected flux which
hits and joins the lower arcade is shown in Figures 2(d)-2(f) as the 
fieldlines which now lie on top of the original arcade.

To illustrate the downflow and deceleration of the reconnected 
flux tubes, we show a height-time plot of the guide field along the 
$x=y=0$ line in Figure 3.
Here, the upward or downward propagating voids are displayed as 
diagonal white streaks of concentrated guide field. 
The first pair of up- and down-flowing streaks,
starting at $tv_{A0}/R=0$, are the tracks of the two initial 
reconnection voids. The upflowing track rapidly hits the upper
Y-line and decelerates, while the downflowing track starts
off at about $v_{A0}/2$, accelerates
as it passes the current sheet midpoint, and then also rapidly
decelerates as it hits the lower Y-line. Guidoni \& Longcope (2007)
suggest that a reconnected flux should accelerate near
$z=0$ as seen here because the magnetic field strength, and thus the 
accelerating Lorentz force, peaks there. 

After the reconnection is turned off at $tv_{A0}/L=0.6$, 
no new tracks appear at the height 
of the initial reconnection site until about $tv_{A0}/L=1.3$.
Then patchy reconnection spontaneously sets in, as several
reconnection outflow pairs appear in rapid sequence.
This is supported by the more detailed views of Figures 1 and 2.
Figures 1(f)-1(h) show the corresponding newly reconnected flux 
tube cross sections flowing up and down at these times. Meanwhile,
Figures 2(c)-2(e) show the continual generation of newly reconnected
fieldlines high up in the current sheet, well after the initial
reconnection has been turned off and those reconnected fields
have retracted. The source of this secondary patchy reconnection may
be numerical resistivity due to the finite grid-scale of the simulation, 
in tandem with perturbations excited by the initial reconnection event, 
and will be explored in a future paper. Isobe et al.\ (2005) found a similar
spontaneous patchy reconnection, but in their case it was due to an
interchange instability, which cannot occur here. Interestingly, 
Figure 1 also shows the reconnected fieldlines spreading in the $\yhat$ 
direction along the current sheet, also in a patchy fashion, instead 
of remaining near the initial reconnection site.  This is in contrast with 
the findings of Ugai et al.\ (2005) for a Harris current sheet without 
guide field. Their results show that initially localized reconnection does 
not expand significantly in this direction. From these preliminary results,
it appears that the guide field can play an important role in 
expanding the reconnection region along the current sheet.

\begin{figure}[t]
\centerline{\includegraphics[width=8.0cm,clip]{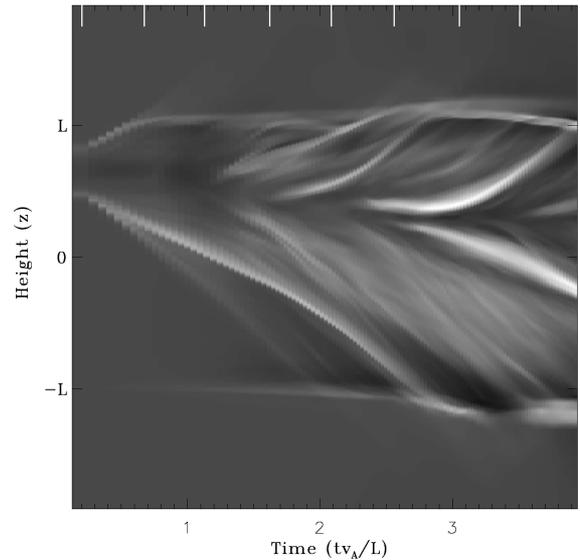}}
\caption{Height-time plot of the guide field at $x=y=0$. The white
tickmarks at the top mark the times of the vector plots of Figure 1.}
\label{fig:htime}
\end{figure}

\section{Conclusions}

We have studied the effect of a brief, localized burst of 
reconnection in a Y-type current sheet with guide field.
The up- and down-flowing flux this creates decelerates
rapidly when it hits the Y-lines and the arcade fields beyond them.
This gives strong support to the theory that the downflowing post-CME
voids observed by TRACE, Yohkoh, and Hinode are in fact downflowing
reconnected magnetic flux loops. This model gives a clear
mechanism to explain why downflowing post-CME voids decelerate rapidly 
when they reach the post-flare arcades, as observed by Sheeley et al.\ 
(2004).  We also find that the velocities of the voids change as they 
propagate through regions of different magnetic field strength in
the current sheet. Finally, we find that the perturbation created
by the initial burst of reconnection is sufficient to excite
spontaneous patchy reconnection events in the current sheet.

\acknowledgments{This work was funded by NASA and 
ONR, with a grant of computer time from 
from the DoD High Performance Computing program.}



\email{M. Linton (e-mail: linton@nrl.navy.mil)}
\label{finalpage}
\lastpagesettings
\end{document}